\documentclass[12pt]{article}
\usepackage{amssymb,amsmath,epsfig}
\allowdisplaybreaks

\begin{document}

\title{\bf Stability of Gravastars with Exterior Regular Black Holes}
\author{M. Sharif \thanks {msharif.math@pu.edu.pk} and Faisal Javed
\thanks{faisalrandawa@hotmail.com}\\
Department of Mathematics, University of the Punjab,\\
Quaid-e-Azam Campus, Lahore-54590, Pakistan.}

\date{}
\maketitle

\begin{abstract}
This paper examines the stability of thin-shell gravastars in the
context of regular spacetimes (Bardeen and Bardeen-de Sitter black
holes). We apply cut and paste approach to construct gravastars
through the matching of interior non-singular de Sitter geometry
with exterior regular black hole. This model contains three regions,
i.e., interior, thin-shell and exterior. The interior and exterior
regions are connected at thin-shell. We investigate physical
viability of the developed model by the energy conditions and
explore its stability by using radial perturbation about the
equilibrium shell radius. It is found that thin-shell gravastars
show large stable regions for the Bardeen-de Sitter black hole as
compared to the Bardeen black hole. It is concluded that stable
regions exist near the formation of expected event horizon.
\end{abstract}

\section{Introduction}

The final outcome of the gravitational collapse of massive objects
containing singularity at their center are referred to as black
holes (BHs). These compact objects are surrounded by a boundary from
which nothing can escape, not even light known as the event horizon.
It is a one-way membrane that allows only to move inside the BH and
acts as a barrier between interior and exterior geometries. Mazur
and Mottola \cite{1} proposed a new image of the collapse by
extending the concept of Bose-Einstein condensation to gravitational
systems known as gravitational vacuum star (gravastar) \cite{2}.
They used cut and paste method to obtain the geometrical structure
of thin-shell gravastars. These geometries are interesting because
they could address two basic issues, one being the challenge of
singularity while another is an information loss paradox related to
BH spacetimes. Such a geometrical structure does not contain the
central singularity and event horizon.

Gravastar has de Sitter geometry as an interior spacetime while the
usual BH as an exterior spacetime like Schwarzschild BH. The
interior and exterior regions are partitioned through a thin layer
of matter surface known as thin-shell. The following geometrical
structure can be characterized into three regions with different
equations of state (EoS). Mathematically, these EoS can be defined
as
\begin{itemize}
\item For interior region ($0\leq r<r_1$), $p=-\sigma$.
\item For thin-shell ($r_1< r<r_2$), $p=\sigma$.
\item For exterior region ($r_2<r$), $p=0=\sigma$.
\end{itemize}
Here, $p$ is the surface pressure and $\sigma$ is the surface energy
density while $r_1-r_2$ represents thickness of the shell. The
presence of matter distribution has great importance to maintain the
stable configuration of a thin-shell that produces enough pressure
to counterbalance the effect of gravitational force. The
characteristics of matter distribution can be determined by using
Israel formalism \cite{3}. The cut and paste approach eliminates the
singularity and event horizon in the geometrical structure of
gravastars \cite{3a}. This technique has also been applied to
construct thin-shell wormholes from different BHs \cite{4}.

Many researchers have studied the new image of gravastar through
various approaches. Visser and Wiltshire \cite{5} developed
thin-shell gravastars from the joining of interior and exterior
spacetimes using cut and paste approach. They also investigated the
stable structure through radial perturbation for some specific EoS.
Carter \cite{6} studied the stability of thin-shell gravastars by
using EoS with different exterior geometries. Bil$\acute{i}$c et al.
\cite{7} introduced a new type of gravastars by replacing de Sitter
interior geometry with Born-Infled phantom. Horvat et al. \cite{8}
extended the concept of gravastars by considering
Reissner-Nordstr\"{o}m spacetime as an exterior geometry. Usmani et
al. \cite{9} also proposed charged gravastars and studied the
entropy of the system. Banerjee et al. \cite{10} introduced an
alternative of braneworld BHs as braneworld gravastars and also
explored their physical characteristics.

G$\acute{a}$sp$\acute{a}$r and R$\acute{a}$cz \cite{11} observed the
stability of gravastars through the inelastic collision of their
surface layer with a dust shell. Horvat et al. \cite{12} considered
the gravastar with continuous pressure and examined stability
through the conventional Chandrasekhar approach. Lobo and Garattini
\cite{13} found the exact solutions of gravastars in noncommutative
geometry and studied their physical characteristics. They explored
the dynamical stability of the transition layer for some specific
cases and found that stable regions are enhanced near the formation
of the expected event horizon. Lobo et al. \cite{14} explored the
stability of gravastars related to the matter distribution in the
transition layer. \"{O}vg\"{u}n et al. \cite{15} constructed
thin-shell gravastars in the background of noncommutative geometry
and examined stability regions through radial perturbation about the
equilibrium shell radius. Shamir and Ahmad \cite{15a} discussed
various physical characteristics like, entropy, the EoS parameter,
length of the shell, energy-thickness relation of the gravastar
shell model in $f(G,T)$ gravity. Yousaf et al. \cite{15b} examined
the stable regions of gravastar and its characteristics in the
background of $f(R,T)$ gravity. Sharif and Waseem \cite{15c}
explored the charged gravastars with conformal motion in $f(R,T)$
gravity.

A regular BH is an outcome of multiple attempts to establish a
feasible interior structure by avoiding the singularity. Bardeen
\cite{17} was the pioneer to introduce exact solutions of the field
equations that contain event horizon with regular center. Later,
some other models of regular BHs were proposed \cite{18}. Moreno and
Sarbach \cite{18a} investigated the dynamical stability of regular
BHs with respect to arbitrary linear fluctuations of the metric and
the electromagnetic field. Zhou et al. \cite{18b} observed the
behavior of the effective potential for the particles and photons in
the spacetime of Bardeen BH. Fernando \cite{19} studied the Bardeen
BH in de Sitter and anti-de Sitter spacetimes. Recently, Li et al.
\cite{19a} examined thermodynamical stability of Bardeen BH through
heat capacity as well as Gibbs free energy and also discussed
thermodynamics of Bardeen-AdS BH.

The cosmological constant ($\Lambda$) is an important parameter to
investigate thin-shell stability. Eiroa and Romero \cite{19b}
examined the stability of thin-shell wormholes constructed from
different BHs. They found stable static solutions with Chaplygin gas
model for different values of charge as well as $\Lambda$. Lobo and
Crawford \cite{19c} studied stability of spherical thin-shell
wormholes in the presence of $\Lambda$ and found that stable regions
are enhanced for large positive value of $\Lambda$. We also analyzed
the linearized stability of thin-shell wormholes developed from
Bardeen and Bardeen-de Sitter BHs with variable EoS \cite{19d}. It
is found that thin-shell becomes more stable in the presence of
cosmological constant.

Regular BHs motivate to develop thin-shell gravastars by considering
regular BHs as an exterior geometry. In this paper, we are
interested to examine the stable characteristics of regular
thin-shell gravastars through radial perturbation. The paper has the
following format. Section \textbf{2} develops the general formalism
of thin-shell gravastars through cut and paste approach. Section
\textbf{3} explains the stability procedure of thin-shell through
radial perturbation about the equilibrium shell radius. We also
observe the corresponding stable regions of thin-shell gravastars.
In the last section, we summarize our results.

\section{Exterior of Gravastars: Regular Black Holes}

The line element of Bardeen-de Sitter BH can be expressed as
\cite{19}
\begin{eqnarray}\nonumber
ds^2&=&-\left(1-\frac{2r^2m}{(r^2+Q^2)^\frac{3}{2}} -\frac{\Lambda
r^2}{3}\right)dt^2+\left(1-\frac{2r^2m}{(r^2+Q^2)^\frac{3}{2}}
-\frac{\Lambda r^2}{3}\right)^{-1}dr^2
\\\nonumber&+&r^2(d\theta^2+\sin^2\theta d\phi^2),
\end{eqnarray}
where $\Lambda$, $Q$ and $m$ denote the cosmological constant,
charge and total mass of the BH. This spacetime can be reduced in
different BH geometries such that
\begin{itemize}
\item If $\Lambda=0$ and $Q\neq0$, then it represents
the Bardeen BH \cite{17}.
\item If $\Lambda=0=Q$, it corresponds to the Schwarzschild BH.
\end{itemize}
The event horizon ($r_h$) of a BH geometry is a point at which the
metric function vanishes.

\subsection{Geometrical Construction of Gravastars}

Here, we briefly discuss the mathematical procedure to develop the
geometry of gravastars. We consider non-singular de Sitter geometry
as an interior metric and regular BHs as an exterior. The
corresponding interior (-) and exterior (+) geometries are expressed
by the line element
\begin{equation}\label{1}
ds^2_{\pm}=-\Phi_{\pm}(r_{\pm})dt_{\pm}^2+\Phi_{\pm}^{-1}(r_{\pm})dr_{\pm}^2
+r_{\pm}^2(d\theta_{\pm}^2+\sin^2\theta_{\pm} d\phi_{\pm}^2),
\end{equation}
where
\begin{equation}\nonumber
\Phi_{-}(r_-)=\left(1-\frac{r^2_-}{\alpha^2}\right),\quad
\Phi_{+}(r_+)=\left(1-\frac{2r^2_+m}{(r^2_++Q^2)^\frac{3}{2}}
-\frac{\Lambda r^2_+}{3}\right),
\end{equation}
and $\alpha$ is a nonzero constant. Visser introduced a well-known
approach to develop thin-shell gravastars by the junction of both
spacetimes that eliminate the event horizon and singularity. For
this purpose, we consider a subset ($\Upsilon^\pm$) of these
manifolds $(\Pi^\pm)$ through cut and paste technique that does not
contain any type of event horizon as well as singularity, i.e.,
$\Upsilon^\pm\subset\Pi^\pm$. Here, $\Upsilon^\pm=\{x^\nu|r_\pm\geq
y(\tau)>r_h\}$, where $x^\nu$, $\tau$ and $y(\tau)$ represent
coordinates of the manifold, proper time on the shell and shell
radius. These subsets $\Upsilon^\pm$ are glued at their common
timelike hypersurface $\partial \Upsilon$, i.e., $\partial
\Upsilon\subset\Upsilon^\pm$. The matching between $\Upsilon^+$ and
$\Upsilon^-$ at throat radius provides a connection between interior
and exterior spacetimes ($\partial\Upsilon\equiv\Upsilon^+\cup
\Upsilon^- $) that follow the radial flare-out condition. This
manifold ($\partial\Upsilon$) represents a thin-shell gravastar
which is geodesically complete.

The corresponding line element of the induced metric at
$\partial\Upsilon$ is given in the following form
\begin{equation}\nonumber
ds^2=-d\tau^2+y(\tau)^2d\theta^2+y(\tau)^2\sin^2\theta d\phi^2,
\end{equation}
and the components of unit normals at $\Upsilon_{\pm}$ can be
expressed as
\begin{equation}\nonumber
n_{\pm}^{\mu}=\left(\frac{\dot{y}}{\Phi_\pm(y)},\sqrt{\Phi_\pm(y)
+\dot{y}^2},0,0\right),
\end{equation}
where $\dot{y}=dy/d\tau$. The components of extrinsic curvature are
defined as
\begin{eqnarray}\nonumber
K_{ij}^{\pm}=-n_{\mu}^{\pm}\left(\Gamma^\mu_{\alpha\beta}\frac{
dx^{\alpha}_{\pm}}{d\eta^{i}}\frac{ dx^{\beta}_{\pm}}{
d\eta^{j}}+\frac{d^2x_{\pm}^\mu}{d\eta^i d\eta^j}\right),\quad
\alpha,\beta=0,1,2,3,\quad i,j=0,2,3,
\end{eqnarray}
and hence
\begin{equation}\label{2}
K_{\tau}^{\tau\pm}=\frac{\Phi'_\pm(y)+2\ddot{y}}{\sqrt{\Phi_\pm(y)+\dot{y}^2}},
\quad K_{\theta}^{\theta\pm}= \frac{\sqrt{
\Phi_\pm(y)+\dot{y}^2}}{y},\quad K^{\phi\pm}_{\phi}=\sin^2\theta
K_{\theta}^{\theta\pm},
\end{equation}
where $\Phi'_\pm(y)=\frac{d\Phi_\pm(y)}{dy}$.

The presence of matter surface produces extrinsic curvature
discontinuity at the hypersurface and its existence can be evaluted
by using Israel formalism. Mathematically, such a matter surface can
be observed, if $(K_{ij}^{+}-K_{ij}^{-}\neq0)$. The characteristics
of matter surface located at thin-shell are determined by the field
equations for the hypersurface referred to as Lanczos equations
\begin{equation}\label{3}
S^{i}_{j}=-\frac{1}{8\pi}\{[K^{i}_{j}]-\delta^{i}_{j}K\},
\end{equation}
where $S^{i}_{j}$ denotes the energy-momentum tensor for
$\partial\Upsilon$, $[K^{i}_{j}]=K^{+i}_{j}-K^{-i}_{j}$ and
$K=tr[K_{ij}]=[K^{i}_{j}]$.  For perfect fluid distribution, the
stress-energy tensor yields
\begin{equation}\nonumber
S^{i}_{j}=\left(\sigma+p\right)u^i u_j+p\delta^{i}_{j},
\end{equation}
here $u_i$, $\sigma$ and $p$ denote the components of shell's
velocity, surface energy density and pressure of the matter surface.
The corresponding $\sigma$ and $p$ of thin-shell gravastars can be
evaluated through the Lanczos equations as
\begin{eqnarray}\label{4}
\sigma&=&-\frac{[K^\theta_\theta]}{4\pi}=-\frac{1}{4\pi
y}\left\{\sqrt{\dot{y}^2+\Phi_+(y)}-\sqrt{\dot{y}^2+\Phi_-(y)}\right\},
\\\nonumber
p&=&\frac{[K^\theta_\theta]+[K_{\tau}^{\tau}]}{8\pi}=\frac{1}{8\pi
y}\left\{\frac{2\dot{y}^2+2y\ddot{y}
+2\Phi_+(y)+y\Phi'_+(y)}{\sqrt{\dot{y}^2+\Phi_+(y)}}\right.\\\label{5}&
-&\left.\frac{2\dot{y}^2+2y\ddot{y} +2\Phi_-(y)+y
\Phi'_-(y)}{\sqrt{\dot{y}^2+\Phi_-(y)}}\right\},
\end{eqnarray}
while
\begin{eqnarray}\label{6}
\sigma+2p&=&\frac{1}{4\pi}[K_{\tau}^{\tau}]=\frac{1}{4\pi}\left\{
\frac{\Phi'_+(y)+2\ddot{y}}{\sqrt{\Phi_+(y)+\dot{y}^2}}-\frac{\Phi'_-(y)+2
\ddot{y}}{\sqrt{\Phi_-(y)+\dot{y}^2}}\right\}.
\end{eqnarray}
Here, we assume that shell's movement along the radial direction
vanishes at equilibrium throat radius say $y_0$, i.e.,
$\dot{y}_0=0=\ddot{y}_0$. Hence, the above equations can be
expressed as
\begin{eqnarray}\label{7}
\sigma_0&=&-\frac{1}{4\pi
y_0}\left\{\sqrt{\Phi_+(y_0)}-\sqrt{\Phi_-(y_0)}\right\},
\\\label{8}
p_0&=&\frac{1}{8\pi y_0}\left\{\frac{2\Phi_+(y_0)+y_0\Phi'_+(y_0)}
{\sqrt{\Phi_+(y_0)}}-\frac{2\Phi_-(y_0)+y_0
\Phi'_-(y_0)}{\sqrt{\Phi_-(y_0)}}\right\},
\end{eqnarray}
and
\begin{equation}\label{9}
\sigma_0+2p_0=\frac{1}{4\pi}\left\{
\frac{\Phi'_+(y_0)}{\sqrt{\Phi_+(y_0)}}-\frac{\Phi'_-(y_0)}
{\sqrt{\Phi_-(y_0)}} \right\},
\end{equation}
where $\sigma_0$ and $p_0$ are the surface energy density and
pressure at $y=y_{0}$, respectively.

\subsection{Energy Conditions and Balance Equation}

To discuss the physical viability of a model, we impose some
geometrical constraints, known as energy conditions. There are four
well-known energy conditions null ($\sigma_0+p_0\geq0$), weak
($\sigma_0\geq0$, $\sigma_0+p_0\geq0$), strong
($\sigma_0+3p_0\geq0$) and dominant ($\sigma_0\pm p_0\geq0$,
$\sigma_0\geq0$). These constraints must be satisfied for normal
matter distribution. These energy conditions are satisfied for
thin-shell gravastars for some specific conditions
given as\\\\
If $\sqrt{\frac{\Phi_+(y_0)}{\Phi_-(y_0)}}\leq1$, then
$\sigma_0\geq0$.\\\\
If
$\sqrt{\frac{\Phi_+(y_0)}{\Phi_-(y_0)}}\leq\frac{\Phi_+(y_0)'}{\Phi_-(y_0)'}$,
then $\sigma_0+p_0\geq0$.\\\\
If
$\sqrt{\frac{\Phi_+(y_0)}{\Phi_-(y_0)}}\leq\frac{4\Phi_+(y_0)+y_0\Phi_+(y_0)'}
{4\Phi_-(y_0)+y_0\Phi_-(y_0)'}$, then $\sigma_0-p_0\geq0$.\\\\
If
$\sqrt{\frac{\Phi_+(y_0)}{\Phi_-(y_0)}}\leq\frac{4\Phi_+(y_0)+3y_0\Phi_+(y_0)'}
{4\Phi_-(y_0)+3y_0\Phi_-(y_0)'}$, then $\sigma_0+3p_0\geq0$.\\

Now, we consider the equation that describes characteristics of
radial pressure in terms of the total energy-momentum tensor
($T_{\alpha\beta}^{\text{total}}$) at the hypersurface as \cite{20}
\begin{equation}\nonumber
[T_{\alpha\beta}^ {\text{total}} n^{\alpha}n^{\beta}] =
\frac{1}{2}S^i_j (K^{i+}_j+K^{i-}_j),
\end{equation}
where the square brackets represent the discontinuity across the
shell. By considering the values of extrinsic curvature for interior
and exterior spacetimes, we obtain the pressure balance equation as
\begin{equation}\label{10}
\Delta^-(y_0)-\Delta^+(y_0)=\frac{\sigma_0}{2}\left(\frac{\Phi_+(y_0)'}{\sqrt{\Phi_+
(y_0)}}+\frac{\Phi_-(y_0)'}{\sqrt{\Phi_-(y_0)}}\right)+\frac{p_0}{2y_0}
(\sqrt{\Phi_+(y_0)}+\sqrt{\Phi_-(y_0)}).
\end{equation}
where $-\Delta^+(y_0)=T_{\alpha\beta}^ {\text{total}}
n^{\alpha}n^{\beta}$ represents the radial tension acting on the
shell. This equation gives the difference between radial tension of
interior and exterior geometries in terms of $\sigma_0$ and $p_0$.
It is noted that for the exterior vacuum solution $\Delta^+(y_0)=0$.
Here, we consider a particular case, $\sigma=0$ for which
Eq.(\ref{10}) reduces to
\begin{equation}\label{11}
\Delta^-(y_0)=\frac{p_0}{2y_0}
\left(\sqrt{\Phi_+(y_0)}+\sqrt{\Phi_-(y_0)}\right).
\end{equation}
This equation directly relates the interior radial tension
$\Delta^-(y_0)$ with a surface pressure of the matter surface
located at the shell. If $\Delta^-(y_0)>0$, then $p_0>0$, that
prevents the geometrical structure of gravastars from collapsing. If
$\Delta^-(y_0)<0$, then $p_0<0$, that hold the expanding behavior of
the shell. The corresponding expansion and collapse of thin-shell
gravastars are shown in Figure \textbf{1}. Initially, thin-shell
shows expanding behavior ($p_0<0$) and then represents the collapse
($p_0>0$).
\begin{figure}\centering
\epsfig{file=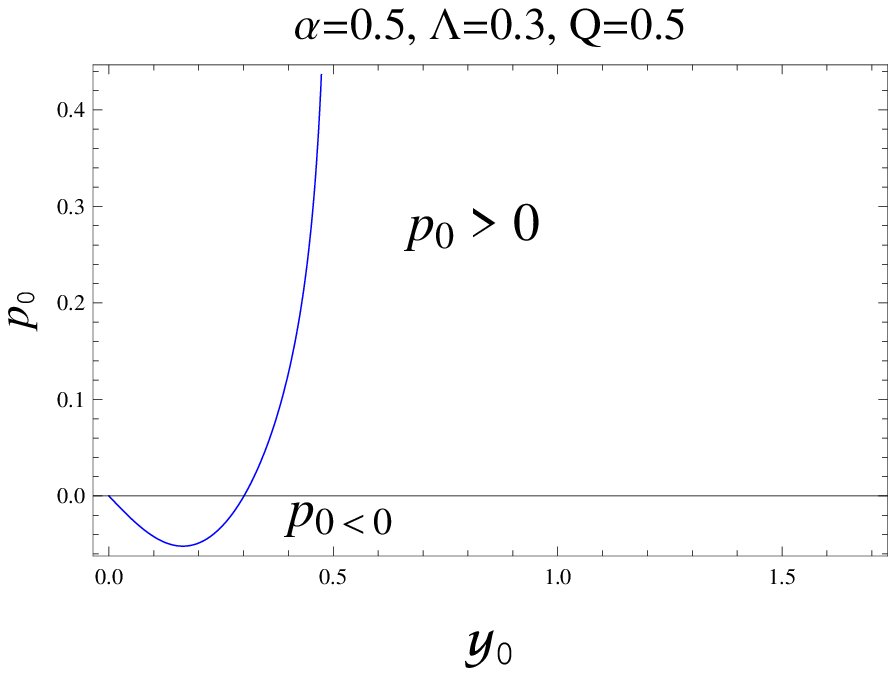,width=.5\linewidth}\epsfig{file=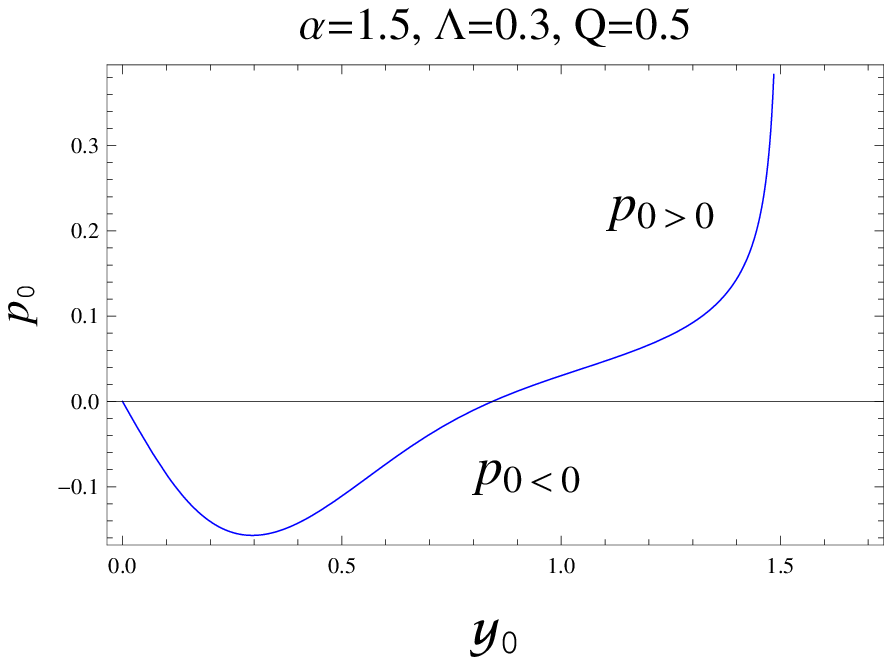,width=.5\linewidth}
\caption{Plots of $p_0$ for different values of $\alpha$. These
plots show the expanding and collapsing behavior of thin-shell
gravastars.}
\end{figure}

\section{Stability Analysis}

In this section, we examine the stability of thin-shell gravastars
through the radial perturbation about the shell's radius at the
equilibrium position. The dynamical characteristics of thin-shell
can be determined through the equation of motion and the
conservation equation. These equations have great importance to
explore the stable regions of a geometrical structure. Firstly, we
derive the equation of motion by rearranging Eq.(\ref{4}) as
\begin{equation}\label{12}
\dot{y}^2+\Omega(y)=0,
\end{equation}
where $\Omega(y)$ is the effective potential that can be expressed
as
\begin{equation}\label{13}
\Omega(y)=\frac{1}{2}\left(\Phi_-(y)+\Phi_+(y)\right)-\frac
{(\Phi_-(y)-\Phi_+(y))^2}{64 \pi ^2 y^2 \sigma ^2}-4 \pi ^2
y^2\sigma ^2.
\end{equation}
For Bardeen-de Sitter and Bardeen BHs, the corresponding effective
potential become
\begin{eqnarray}\nonumber
\Omega_{BDS}(y)&=&1-4 \pi ^2 y^2 \sigma ^2-\frac{1}{64 \pi ^2 y^2
\sigma ^2}\left(\frac{y^2}{\alpha ^2}-\frac{1}{3} y^2
\left(\Lambda+\frac{6
m}{\left(Q^2+y^2\right)^{3/2}}\right)\right)^2\\\nonumber&-&\frac{\Lambda
y^2}{6}-\frac{m y^2}{\left(Q^2+y^2\right)^{3/2}}-\frac{y^2}{2 \alpha
^2},\\\nonumber \Omega_B(y)&=& 1-\frac{y^2}{2 \alpha ^2}-\frac{y
m}{\left(y^2+Q^2\right)^{3/2}}-\frac{1}{64 \pi ^2 \sigma
^2}\left(\frac{y}{\alpha ^2}-\frac{2
m}{\left(y^2+Q^2\right)^{3/2}}\right)^2\\\nonumber&-&4 \pi ^2 y^2
\sigma ^2,
\end{eqnarray}
respectively. Also, $\sigma$ and $p$ in terms of potential function
can be written as
\begin{eqnarray}\label{14}
\sigma&=&-\frac{1}{4\pi y}\left\{\sqrt{\Omega(y)+\Phi_+(y)
}-\sqrt{\Omega(y)+\Phi_-(y)}\right\},
\\\label{15}
p&=&\frac{2(\Omega+\Phi_+)+y(\Omega' + \Phi'_+)}{8\pi
y\sqrt{\Omega(y)+\Phi_+(y)}} -\frac{2(\Omega+\Phi_-)+y(\Omega' +
\Phi'_-)}{8\pi y\sqrt{\Omega(y)+\Phi_-(y)}}.
\end{eqnarray}

Secondly, we analyze that $\sigma$ and $p$ follow the conservation
equation
\begin{equation}\nonumber
p \frac{d}{d\tau}(4\pi y^2)+\frac{d}{d\tau}(4\pi y^2\sigma)=0,
\end{equation}
which leads to
\begin{equation}\nonumber
\sigma'=-\frac{2(\sigma+p(\sigma))}{y}.
\end{equation}
To observe the stability, we consider radial perturbation through
Taylor series up to second-order terms. Therefore, we expand the
effective potential about $y=y_0$
\begin{equation}\nonumber
\Omega(y)=\Omega(y_{0})+(y-y_{0})\Omega'(y_{0})+\frac{1}{2}
(y-y_{0})^2\Omega''(y_{0})+O[(y-y_{0})^3].
\end{equation}
It is found that $\Omega(y_0)=0=\Omega'(y_0)$. Consequently, the
above equation turns out to be
\begin{equation}\label{16}
\Omega(y)=\frac{1}{2}(y-y_{0})^2\Omega''(y_{0}).
\end{equation}
As the mass of thin-shell can be expressed as $M(y)=4\pi y^2
\sigma$, the corresponding second derivative of effective potential
at $y=y_0$ in terms of $M(y_0)$ yields
\begin{eqnarray}\nonumber \Omega''(y_0)&=&\frac{2
M(y_0) M'(y_0)}{y^3_0}-\frac{y_0^2 (\Phi_-(y_0)-\Phi_+(y_0))
\left(\Phi_-(y_0)'' -\Phi_+(y_0)''\right)}{2M(y_0)^2}\\\nonumber&
+&\frac{2 y_0^2 (\Phi_-(y_0)-\Phi_+(y_0)) M'(y_0) \left(\Phi_-'(y_0)
-\Phi_+'(y_0)\right)} {M(y_0)^3}-\frac{M'(y_0)^2}
{2y_0^2}\\\nonumber&-&\frac{y_0^2 \left(\Phi_-'(y_0)-\Phi_+'
(y_0)\right)^2}{2M(y_0)^2}-\frac{3 y_0^2 (\Phi_-(y_0)-\Phi_+(y_0))^2
M'(y_0)^2}{2M(y_0)^4}\\\nonumber&-&\frac{2 y_0
(\Phi_-(y_0)-\Phi_+(y_0)) \left(\Phi_-'(y_0)-\Phi_+'(y_0)
\right)}{M(y_0)^2}+\frac{(\Phi_-''(y_0)+\Phi_+''(y_0))}{2}
\\\nonumber&-& \frac{3 M(y_0)^2}{2y_0^4}+\frac{ y_0^2
(\Phi_-(y_0)-\Phi_+(y_0))^2 M''(y_0)}{2M(y_0)^3}-\frac{M(y_0)
M''(y_0)}{2y_0^2}\\\label{17}&
-&\frac{(\Phi_-(y_0)-\Phi_+(y_0))^2}{2M(y_0)^2}+\frac{2 y_0
(\Phi_-(y_0)-\Phi_+(y_0))^2 M'(y_0)}{M(y_0)^3},
\end{eqnarray}
where
\begin{eqnarray}\nonumber M'(y_0)=-8\pi y_0 p_0,\quad
M''(y_0)=-8\pi p_0+16\pi \eta_0^2 (\sigma_0+p_0),
\end{eqnarray}
here $\eta_0^2=dp/d\sigma|_{y=y_0}$ represents the EoS parameter.
The potential function and its second derivative explain the
stability of thin-shell gravastars.
\begin{figure}\centering
\epsfig{file=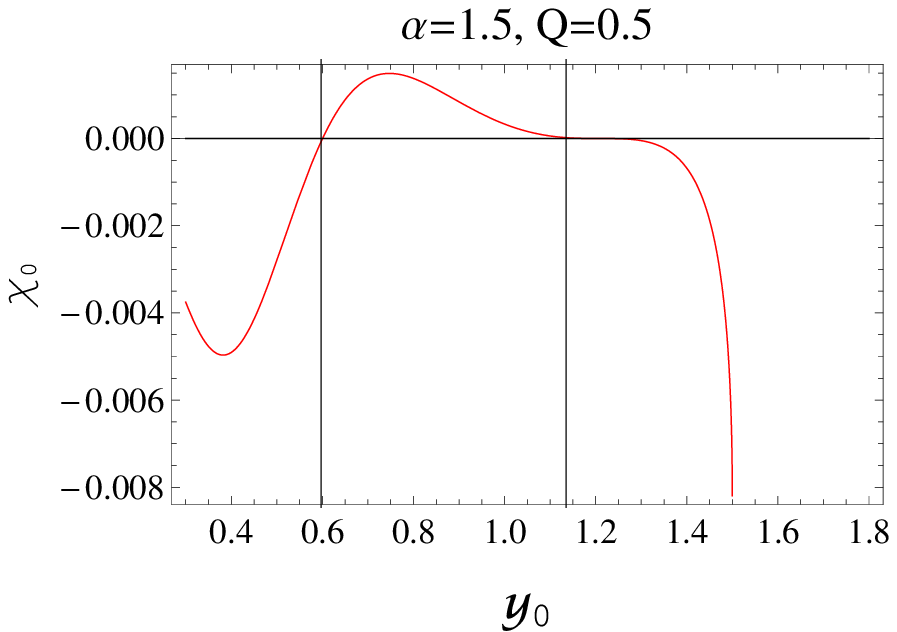,width=.5\linewidth}\epsfig{file=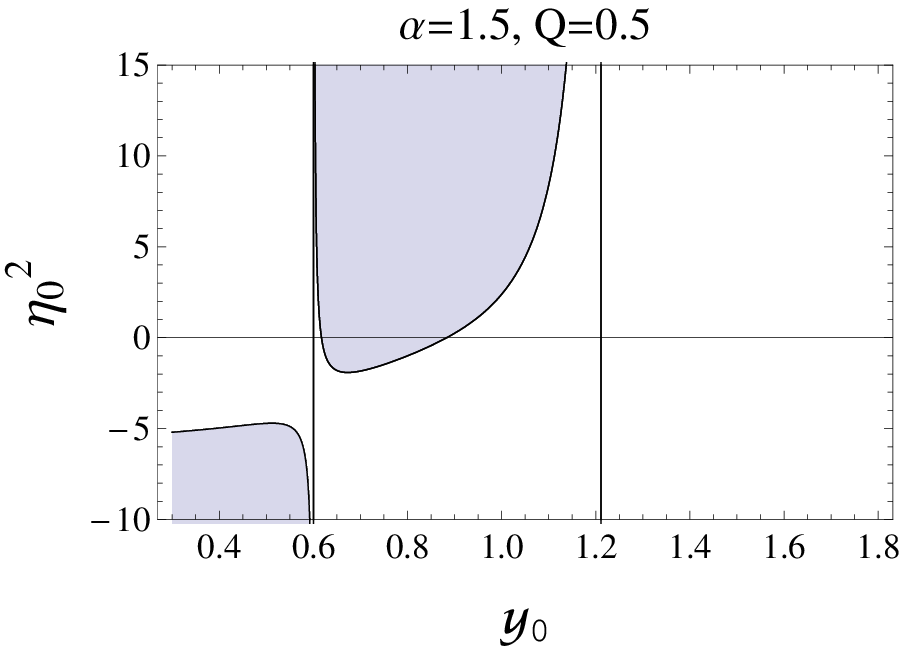,width=.5\linewidth}
\caption{Left and right plots represent the graphical behavior of
$\chi_0$ and $\eta_0^2$ for $\Lambda=0$, respectively. In the
absence of $\Lambda$, these figures explain the characteristics of
thin-shell gravastars for the Bardeen BH. The shaded regions
indicate the stable regions.}
\epsfig{file=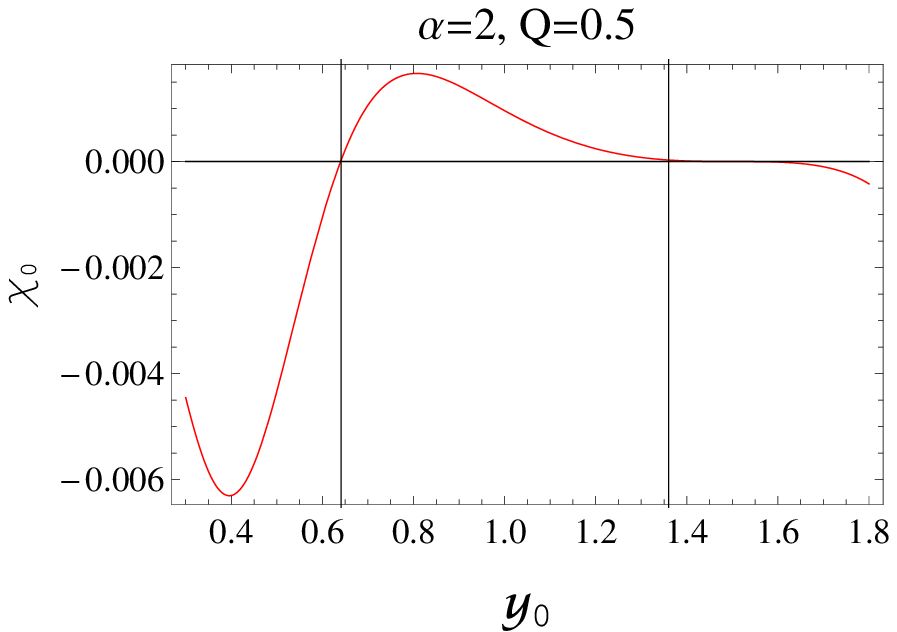,width=.5\linewidth}\epsfig{file=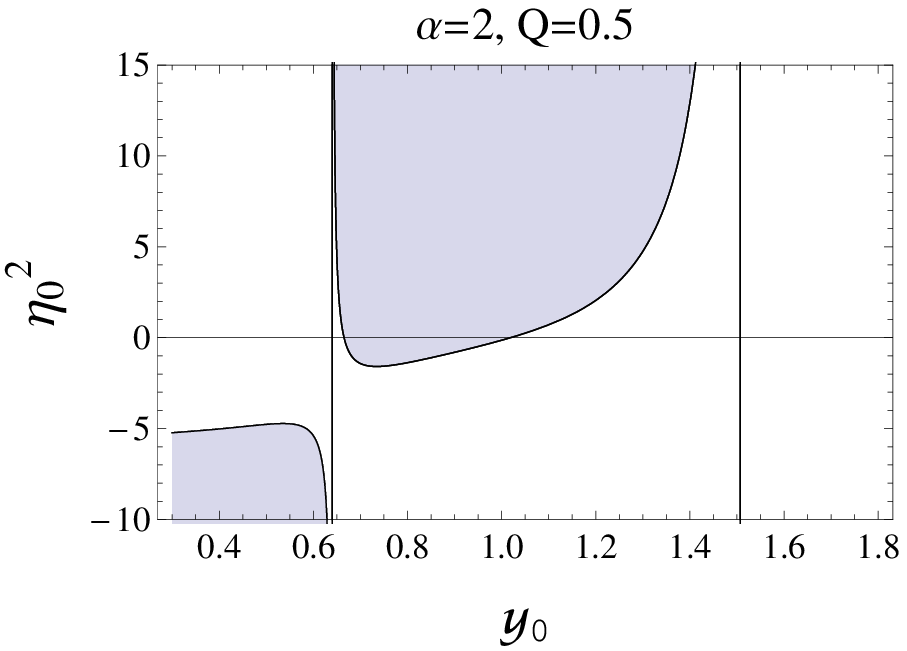,width=.5\linewidth}
\caption{Left and right plots represent the graphical behavior of
$\chi_0$ and $\eta_0^2$ for $\Lambda=0$, respectively.}
\end{figure}

The stable and unstable structures can be characterized as follows
\cite{21}\\\\
(i) \quad If $\Omega''(y_{0})>0$, then
it shows stable behavior.\\\\
(ii) \quad If $\Omega''(y_{0})<0$, it expresses the unstable
behavior.\\\\
(iii) \quad If $\Omega''(y_{0})=0$, then it is unpredictable.\\\\
We are interested in stable configuration of thin-shell gravastars,
i.e., $\Omega''(y_{0})>0$. Therefore, Eq.(\ref{17}) can be expressed
as
\begin{eqnarray}\nonumber
&&\left(-512 \pi ^4 y_0^4 (2 \eta_0^2+3) p_0 \sigma ^5-256 \pi ^4
y_0^4 (4 \eta_0^2+3) \sigma_0 ^6-1024 \pi ^4 y_0^4 p_0^2 \sigma_0
^4\right.\\\nonumber&+&\left.y_0 \sigma_0 \left(y_0 \sigma_0
\left(\Phi_-(y_0)'' \left(16 \pi ^2 y_0^2 \sigma_0 ^2
-\Phi_-(y_0)+\Phi_+(y_0)\right)+\Phi_+(y_0)'' \left(\Phi_-(y_0)
\right.\right.\right.\right.\\\nonumber&+&\left.\left.\left.\left.16
\pi ^2 y_0^2 \sigma_0 ^2-\Phi_+(y_0)\right)\right)+2 \Phi_-(y_0)'
\left(y_0 \sigma_0 \Phi_+(y_0)'-2(2 p_0+\sigma_0 )
\left(\Phi_-(y_0)\right.\right. \right.\right.
\\\nonumber&-& \left.\left.\left.\left.\Phi_+(y_0)\right)\right)-y_0
\sigma_0 \left(\Phi_-(y_0)'\right)^2-y_0 \sigma_0
\left(\Phi_+(y_0)'\right)^2+4(2 p_0+\sigma_0 )
\left(\Phi_-(y_0)\right.\right.\right.\\\nonumber&-&
\left.\left.\left.\Phi_+(y_0)\right) \Phi_+(y_0)' \right)+2 (2
\eta_0^2-5) p_0 \sigma_0 (\Phi_-(y_0)-\Phi_+(y_0))^2-12 p_0^2
\left(\Phi_-(y_0)\right.\right.\\\nonumber&-&\left.\left.\Phi_+(y_0)
\right)^2+(4 \eta_0^2-1) \sigma_0 ^2 (\Phi_-(y_0)-
\Phi_+(y_0))^2\right)(32 \pi ^2 y_0^4 \sigma_0 ^4)^{-1}>0.
\end{eqnarray}
The stable condition of thin-shell can be written in the following
form
\begin{equation}\label{18}
\Omega''(y_0)<0 \quad \Rightarrow \quad
\chi(y_0)\eta_0^2-\textit{A}(y_0)<0.
\end{equation}
Here, $\chi(y_0)=\chi_0$ is the coefficient of EoS parameter
($\eta_0^2$) and $\textit{A}(y_0)=\textit{A}_0$ is the remaining
term of the above expression in which $\eta_0^2$ does not involve.
We discuss the geometrical behavior of thin-shell gravastars through
the stability regions. The stable regions can be characterized as
follows\\\\
(i) \quad If $\chi_0<0$, then
$\eta_0^2<\textit{A}_0/\chi_0$.\\\\
(ii) \quad If $\chi_0>0$, then
$\eta_0^2>\textit{A}_0/\chi_0$,\\\\
where
\begin{eqnarray}\nonumber
A_0&=&-256 \pi ^4 y_0^4 \sigma_0 ^4 \left(4 p_0^2+6 p_0 \sigma_0 +3
\sigma_0 ^2\right)+y_0 \sigma_0  \left(y_0\sigma_0
\left(\Phi_-(y_0)'' \left(16 \pi ^2 y_0^2 \sigma_0
^2\right.\right.\right.\\\nonumber&-&\left.\left.\left.
\Phi_-(y_0)+\Phi_+(y_0)\right)+ \Phi_+(y_0)'' \left(16 \pi ^2 y_0^2
\sigma_0 ^2+\Phi_-(y_0)-\Phi_+(y_0)\right)
\right)\right.\\\nonumber&+&\left.2 \Phi_-(y_0)' \left(y_0 \sigma_0
\Phi_+(y_0)'-2 (\Phi_-(y_0)-\Phi_+(y_0)) (2 p_0+\sigma_0
)\right)\right.\\\nonumber&-&\left.y_0 \sigma_0
\left(\Phi_-(y_0)'\right)^2-y_0 \sigma_0
\left(\Phi_+(y_0)'\right)^2+4\Phi_+(y_0)' (2 p_0+\sigma_0 )
\left(\Phi_-(y_0)\right.\right.\\\nonumber&-&\left.\left.
\Phi_+(y_0) \right)\right)-\Phi_-(y_0)^2 \left(12 p_0^2+10 p_0
\sigma_0 +\sigma_0 ^2\right)+2 \Phi_-(y_0)\left(12 p_0^2+\sigma_0
^2\right.\\\nonumber&+&\left.10 p_0 \sigma_0 \right) \Phi_+(y_0)
-\Phi_+(y_0)^2 \left(12 p_0^2+10 p_0 \sigma_0 +\sigma_0
^2\right),\\\nonumber \chi_0&=&4 \sigma_0 (p_0+\sigma_0 )
\left((\Phi_-(y_0)-\Phi_+(y_0))^2-256 \pi ^4 y_0)^4 \sigma_0
^4\right).
\end{eqnarray}

We explain the stable regions of thin-shell gravastars through the
graphical behavior of $\chi_0$ and $\eta_0^2$ for $m=0.5$. We study
the effect of $\alpha$, charge and cosmological constant on the
geometrical structure of thin-shell gravastars. According to stable
condition, if $\chi_0>0$ then the stable region is the area above
the plots of $A_0/\chi_0$ and vice-versa. Figure \textbf{2} explains
that for the left plot $\chi_0\leq0$ if $y_0\in[0.2,0.6]$ and
$\chi_0\geq0$ if $y_0\in[0.6,1.7]$. Thus the stable regions
represent the area less than $A_0/\chi_0$ if $y_0\in[0.2,0.6]$ and
greater than $A_0/\chi_0$ if $y_0\in[0.6,1.7]$. The point $y_0=0.6$
denotes the event horizon at which $A_0/\chi_0$ becomes infinite. It
is found that near the expected event horizon, stability regions
must exist. Similarly, we observe the stability regions and
corresponding event horizons for different values of $\alpha$, $Q$
and $\Lambda$. Figure \textbf{3} shows stable regions are enhanced
by increasing $\alpha$ for some specific values of $Q=0.5=m$ while
Figure \textbf{4} represents the influence of charge on the
stability of thin-shell gravastars. It is observed that stable
regions are decreased by increasing charge of the exterior geometry.
Figure \textbf{5} indicates that stable regions are greatly affected
by the cosmological constant which shows that stable regions are
enhanced with the cosmological constant.
\begin{figure}\centering
\epsfig{file=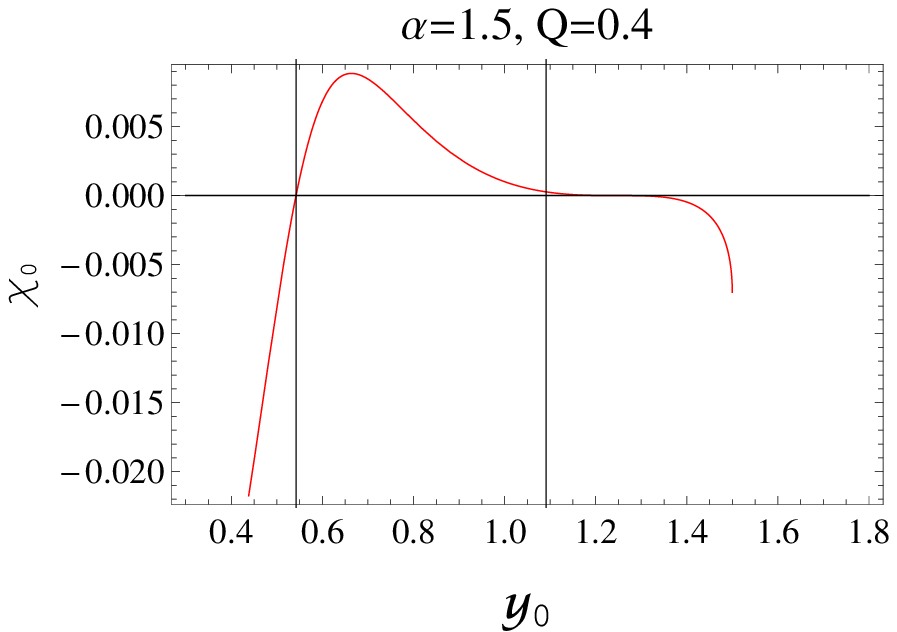,width=.5\linewidth}\epsfig{file=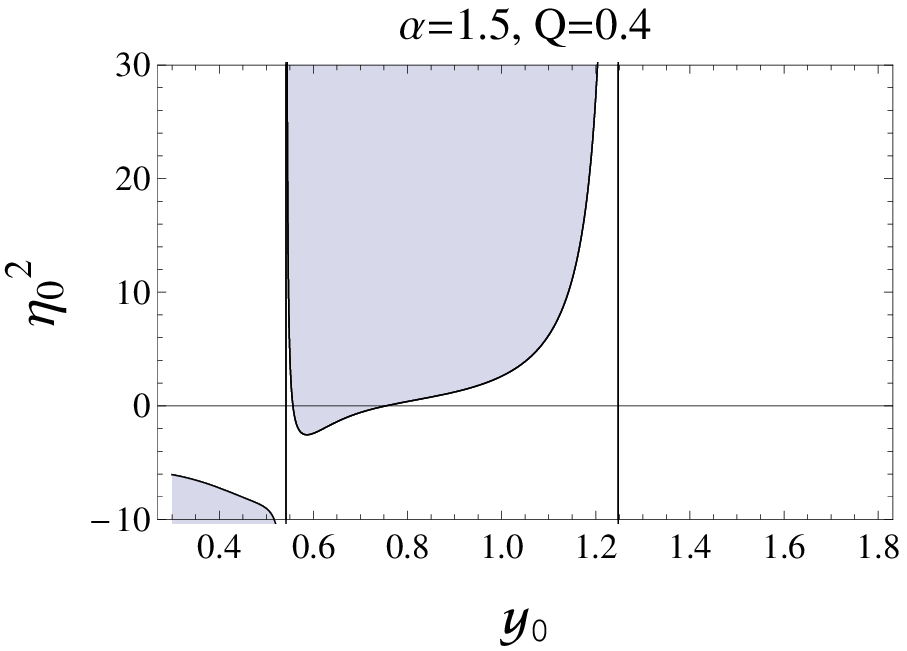,width=.5\linewidth}
\epsfig{file=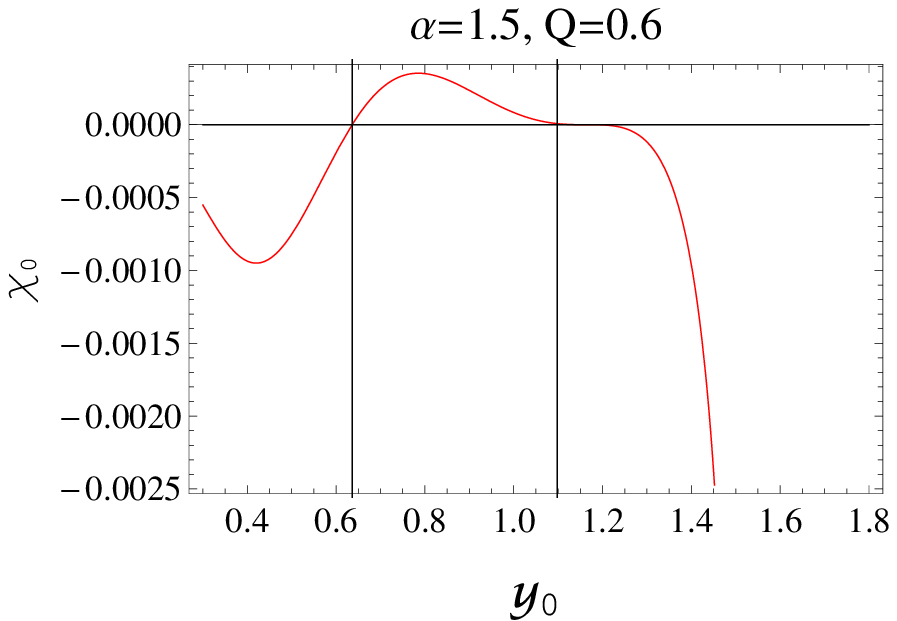,width=.5\linewidth}\epsfig{file=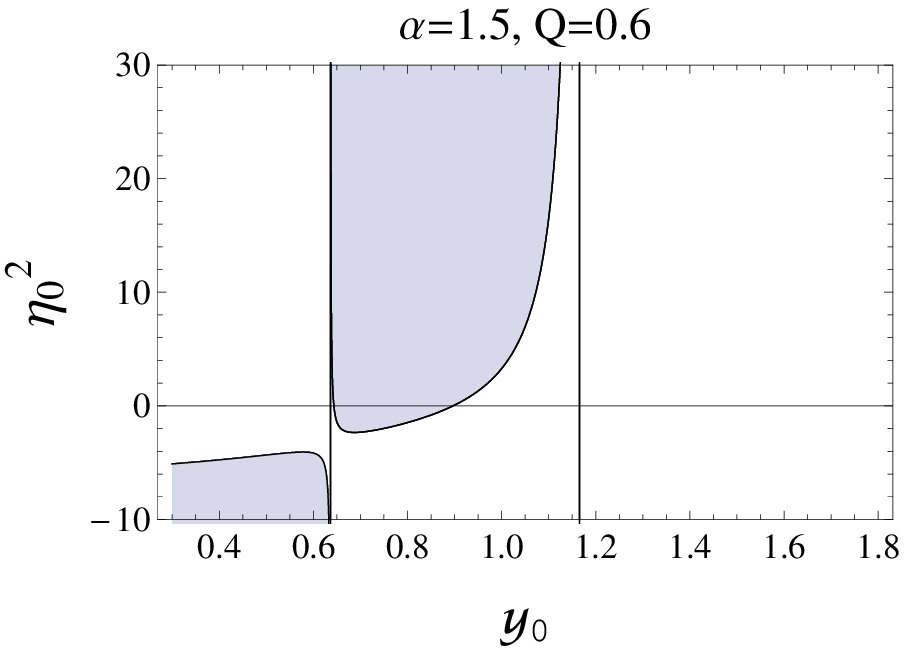,width=.5\linewidth}
\caption{Left and right plots represent the graphical behavior of
$\chi_0$ and $\eta_0^2$ for $\Lambda=0$, respectively.}
\end{figure}
\begin{figure}\centering
\epsfig{file=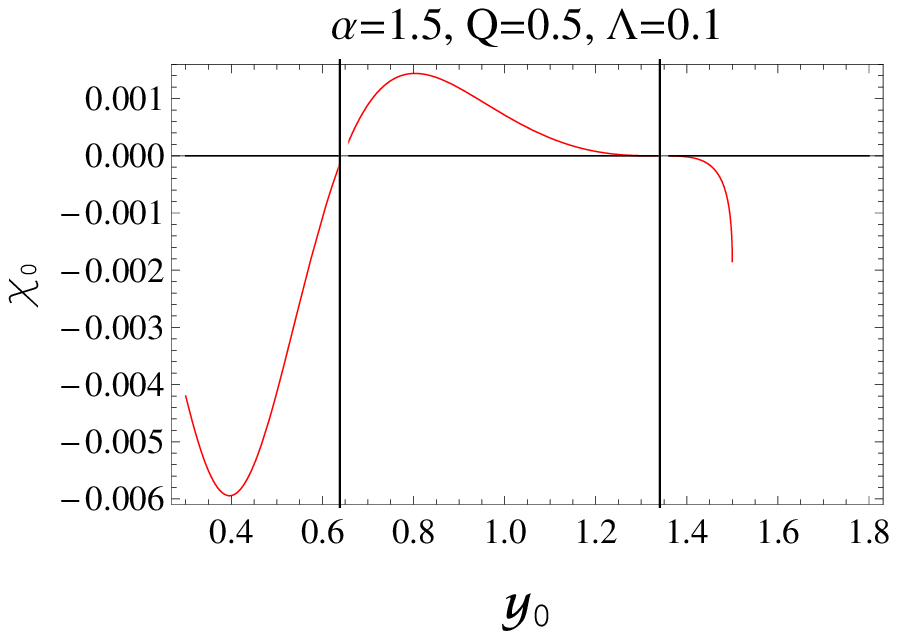,width=.5\linewidth}\epsfig{file=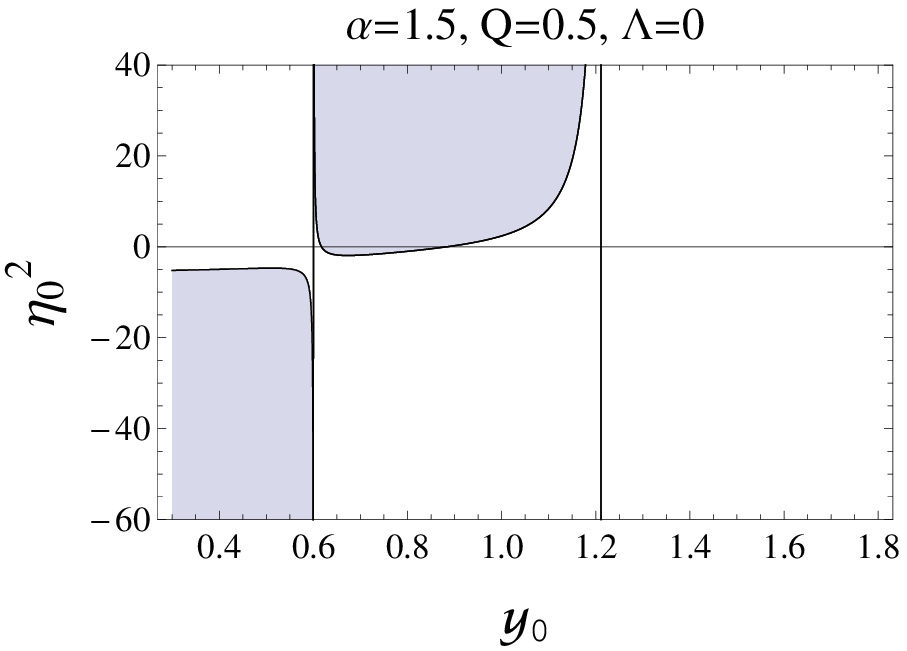,width=.5\linewidth}
\epsfig{file=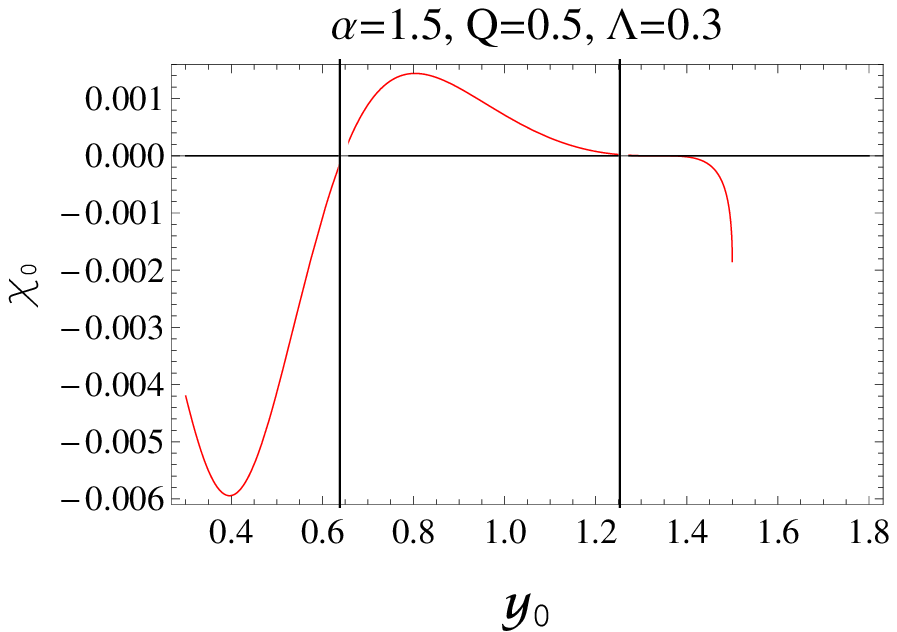,width=.5\linewidth}\epsfig{file=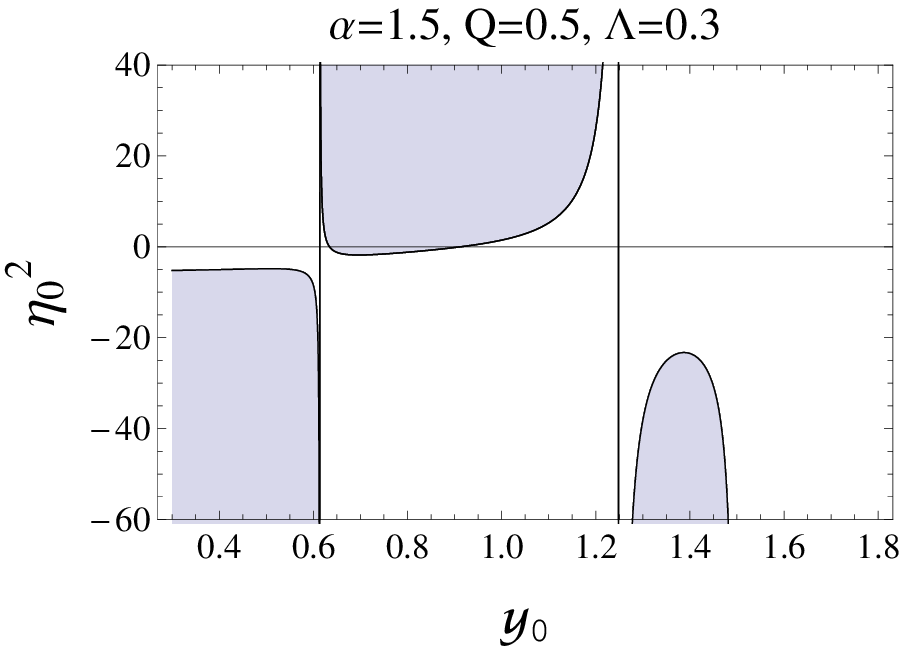,width=.5\linewidth}
\epsfig{file=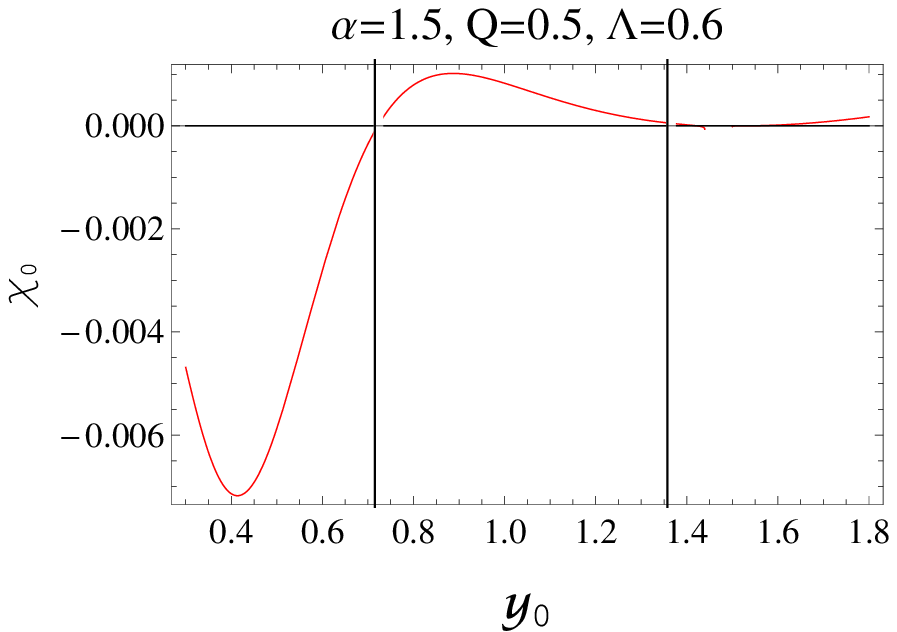,width=.5\linewidth}\epsfig{file=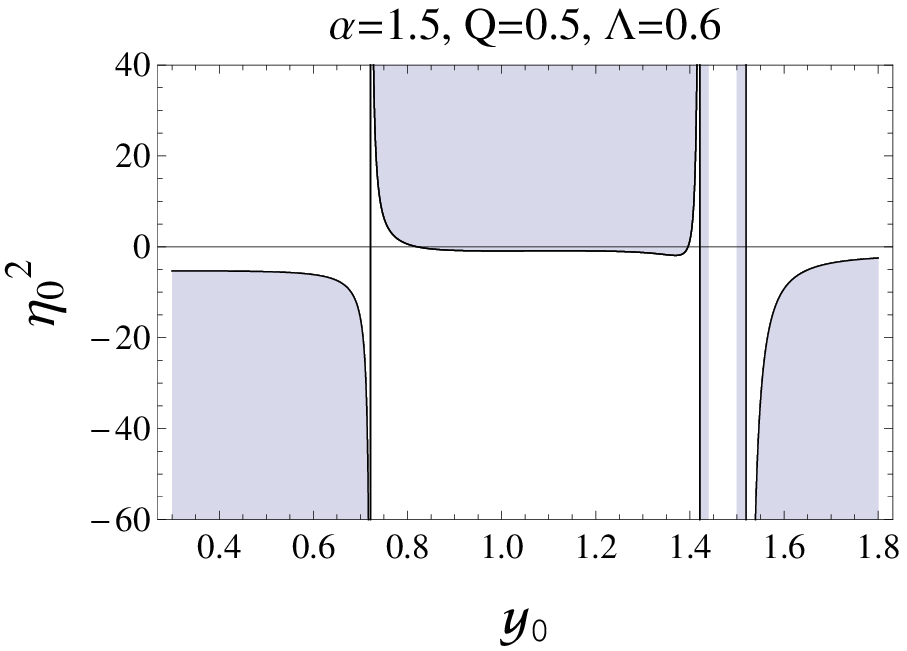,width=.5\linewidth}
\caption{Left and right plots represent the graphical behavior of
$\chi_0$ and $\eta_0^2$, respectively. These plots illustrate the
effect of $\Lambda$ on the stable behavior of thin-shell gravastars
for the Bardeen-de Sitter spacetime.}
\end{figure}

\section{Final Remarks}

In this paper, we have studied the stability of a specific class of
thin-shell gravastars in the background of regular BHs. For this
purpose, we have considered cut and paste approach to develop the
geometrical structure of thin-shell gravastars by the matching of
interior non-singular de Sitter spacetime with exterior regular BH.
The characteristics of matter surface located at thin-shell are
determined by using Israel formalism and the Lanczos equations. For
the viability of the developed structure, we have examined the
energy conditions and found some constraints over the metric
functions of both spacetimes. It is observed that thin-shell
represents expanding and collapsing behavior (Figure \textbf{1}).

We have also studied the stable configurations of thin-shell
gravastars for the Bardeen and Bardeen-de Sitter geometry as
exterior line elements. We have analyzed the stable characteristics
of thin-shell gravastars through the radial perturbation about the
equilibrium shell radius. The stable regions of thin-shell
gravastars are observed through graphs. Figures \textbf{2} and
\textbf{3} indicate that stability of thin-shell gravastars is
increased by increasing $\alpha$. For Bardeen BH, it is found that
stable regions are decreased by increasing charge of the exterior
geometry (Figure \textbf{4}). For Bardeen-de Sitter BH, stable
regions are enhanced with cosmological constant (Figure \textbf{5}).
We would like to mention here that regular BHs as exterior line
element of gravastars show more stable behavior. It is concluded
that thin-shell gravastars are more stable for the Bardeen-de Sitter
BH rather than Bardeen BH.

\section*{Acknowledgement}

We would like to thank the Higher Education Commission, Islamabad,
for its financial support through
\emph{6748/Punjab/NRPU/RD/HEC/2016}.

\end{document}